\documentclass[11pt,oneside]{article}

\usepackage[numbers,sort&compress]{natbib}
\setcitestyle{numbers}

\usepackage{ragged2e}
\usepackage{placeins}
\usepackage{pbox}

\newcommand{\E}{\mathbb{E}}

\usepackage{courier}

\usepackage[hang,flushmargin]{footmisc}

\usepackage{amsmath, amssymb, amsthm, amsfonts, graphicx, mathrsfs, array, stackrel, enumerate, xcolor}
\usepackage{url, overpic, times, eurosym}
\usepackage{booktabs, multirow}
\usepackage{caption,newfloat}
\DeclareCaptionType{InfoBox}

\usepackage{todonotes}
\usepackage{tabularx}
\newcolumntype{Y}{>{\centering\arraybackslash}X}

\usepackage{float}
\restylefloat{table}



\allowdisplaybreaks

\title{Investment sizing with deep learning prediction uncertainties for high-frequency Eurodollar futures trading.} 

\author{Trent Spears\thanks{University of Oxford, Oxford-Man Institute of Quantitative
    Finance.} \thanks{
    Corresponding author.  E-mail: trent@robots.ox.ac.uk}, Stefan Zohren\footnotemark[1], Stephen Roberts\footnotemark[1]
}
\date{July 30, 2020}

\begin{document}
\thispagestyle{plain} \maketitle

\renewcommand{\abstractname}{Summary}
\begin{abstract}
\noindent In this work we show that prediction uncertainty estimates gleaned from deep learning models can be useful inputs for influencing the relative allocation of risk capital across trades.  In this way, consideration of uncertainty is important because it permits the scaling of investment size across trade opportunities in a principled and data-driven way.  We showcase this insight with a prediction model and find clear outperformance based on a Sharpe ratio metric, relative to trading strategies that either do not take uncertainty into account, or that utilize an alternative market-based statistic as a proxy for uncertainty.  Of added novelty is our modelling of high-frequency data at the top level of the Eurodollar Futures limit order book for each trading day of 2018, whereby we predict interest rate curve changes on small time horizons.  We are motivated to study the market for these popularly-traded interest rate derivatives since it is deep and liquid, and contributes to the efficient functioning of global finance -- though there is relatively little by way of its modelling contained in the academic literature. Hence, we verify the utility of prediction models and uncertainty estimates for trading applications in this complex and multi-dimensional asset price space. \\

\noindent {\bf Keywords:} Financial time-series analysis, high-frequency data, interest rate derivatives, deep learning.

\end{abstract}

\section{Introduction}

In practice, it is common for investors to size investment positions based on conviction in trade ideas.  Some traders highlight the importance of investment sizing at least with the view to leverage exposure to the positive right-tail of the profit and loss distribution \cite{Don19}.  However, there is relatively little by way of academic discussion regarding achieving the task in a data-driven way.  Inspired by recent innovations in financial machine learning, we present a case for utilizing uncertainty estimates gleaned from a prediction model, for the purpose of investment sizing within a trading strategy.  Further, we show how this can improve relative investment performance.

\vspace{0.25 \baselineskip}

\noindent
The trading strategy depends on a financial asset price prediction model at its core.  In this context, and in the era of big data, deep learning models often present as the state of the art.  This is particularly true for models estimated given high-frequency limit order book data, whereby many millions of pieces of unique information may be collected and processed according to data-intensive algorithms. In recent work, prediction models for financial time-series have been improved on by deep convolutional neural networks (CNN) and long short-term memory (LSTM) networks \cite{Chen16, Bor19}.  A state-of-the-art CNN-LSTM with inception (CNN-LSTM Inc) was shown to be of utility for formulating financial prediction as a classification problem \cite{Zhang1}.

\begin{center}
\makebox[\textwidth]{
  \includegraphics[height=2.9in, width=6.6in]{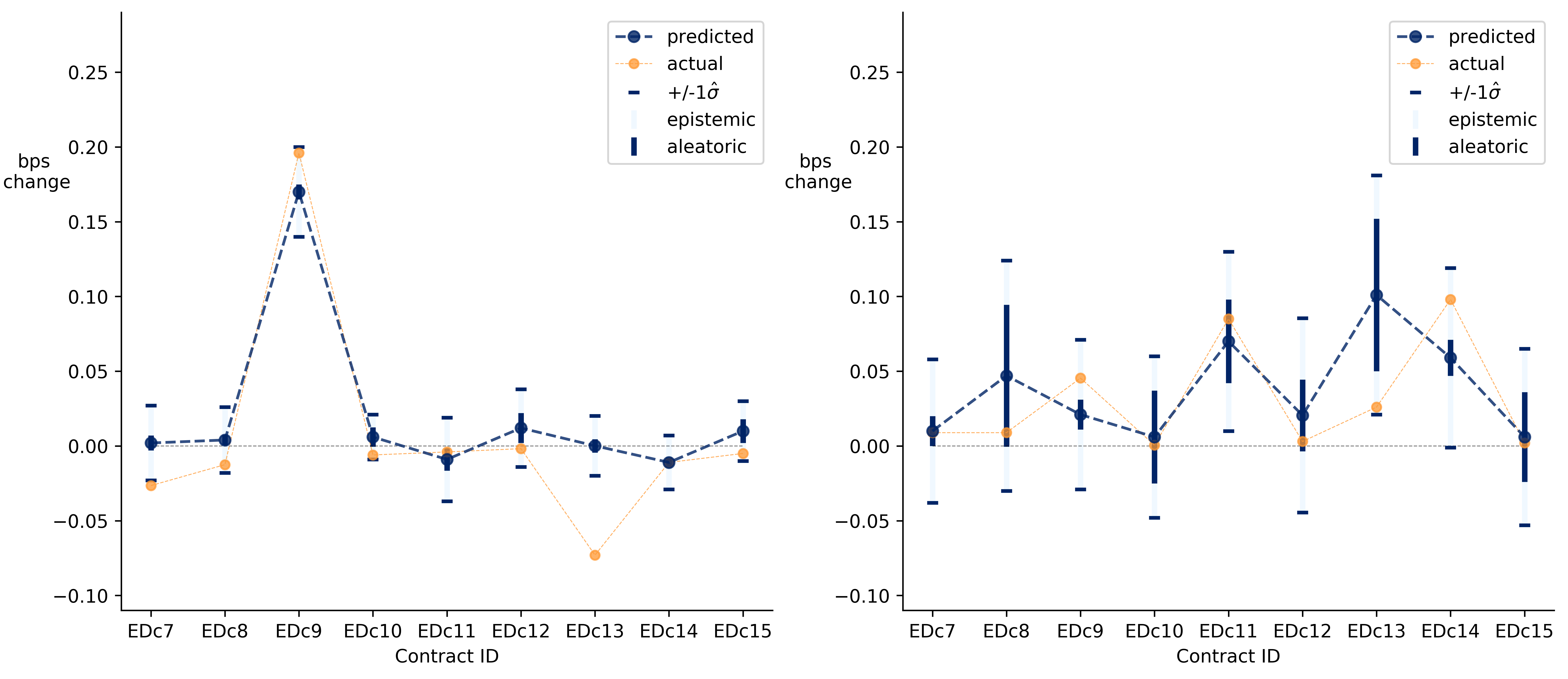}}\par
  \captionof{figure}{We utilize deep learning models of changes in a subset of the interest rate curve for a collection of liquid Eurodollar futures contracts, and also recover uncertainty estimates around predictions.  Above, predictions for the next event time are shown in dark blue, versus the actual realised outcome in orange.  Total uncertainty is shown as the one-standard deviation dashed light blue lines, which subsumes an estimate of aleatoric (epistemic) in the part shaded dark blue (ice blue).  The left-hand image depicts a prediction more certain, and corresponding closer to the target outcome, relative to the right-hand image.
 \label{fg:A}}
\end{center}

\noindent We build from this work to model high-frequency data at Level 1 of the Eurodollar Futures\footnote{Here, we note the class of derivative in focus are the popularly traded interest rate derivatives, whose reference contract interest rate applies to a hypothetical United States dollar-denominated unsecured bank funding deposit -- rather than a foreign exchange derivative on the EURUSD currency pair.} limit order book.  Important variations include our framing of prediction as a regression problem for modelling outright price changes, and that both our input and output data are over a multi-dimensional asset price space.
Further, this approach complements recent literature advancing interest rate derivative price prediction \cite{Kon18, Gon19}.  

\vspace{0.25 \baselineskip}

\noindent 
However, while there is ongoing innovation and success applying deep learning prediction models in finance, less has been shown by way of retrieving meaningful prediction uncertainties.  This is despite innovations and demonstrable utility within a diverse range of application domains, including computer vision, medicine and astrophysics \cite{Ken17, Lei17, Cobb19}.  Of course, prediction uncertainty can arise from a variety of sources, though two stylized facets of uncertainty are often modelled.  On one hand, aleatoric uncertainty; we model this explicitly as expressed through a (spatial) heteroskedastic variance-covariance matrix of the predicted prices.  On the other hand, and less directly, we model epistemic uncertainty as induced by the model specification.  In the context of deep learning, pseudo-Bayesian approximations to epistemic uncertainty have been proposed.  For example, it has been shown that epistemic uncertainty can be cheaply approximated in deep learning models with dropout \cite{Gal16}.  Given that dropout is often a component of deep learning models for financial prediction, and the relative ease by which dropout sampling can be implemented, this approximation seems a practical, yet relatively underexplored, component of prediction uncertainty for financial applications; refer to \cite{Zhang2} for an early attempt.

\vspace{0.25 \baselineskip}

\noindent In any case, the utilization of model predictions and uncertainty estimates -- such as displayed in Figure 1 -- can become apparent in the details of a trading strategy.  A proof of concept of potential usefulness can be found in improving standard and preliminary performance metrics over which trading strategies are typically evaluated.  In particular, we target improvement in the Sharpe ratio \cite{Sharpe94}.  Some authors have improved on a vanilla long/short trading strategy by directly incorporating a measure of Sharpe in the loss function \cite{Choey97, Lim19}, or by applying parameterised decision rules for investment sizing \cite{Tow99}.  Our approach is closer in spirit to the latter: with the addition of reference uncertainty estimates for model-based beliefs of price returns, we can optimize the Sharpe metric inspired by how some successful traders intuitively adjust trade sizes in the real world – by scaling into high conviction trades (corresponding to lower uncertainty) offering sufficiently large reward with relatively more bankroll.

\vspace{0.25 \baselineskip}

\noindent In the subsequent sections, beginning with Section 2, we specify the data set and preprocessing steps (2.1), the model (2.2), and the trading strategy (2.3).  Results and discussion are offered in Section 3.  We conclude in Section 4 and offer avenues for future research.

\begin{table}
\centering
\caption{Median daily number of quotes, trades, and total volume executed for Eurodollar futures contracts EDc1 to EDc15.  Data recorded on each (US) trading day of 2018. Units are in 000s. Data source: Thomson Reuters.  }
\small\addtolength{\tabcolsep}{-4pt}
\begin{tabular}[c]{ c @{\hskip 0.14in} c @{\hskip 0.14in} c @{\hskip 0.14in} c @{\hskip 0.14in} c @{\hskip 0.14in} c @{\hskip 0.14in} c @{\hskip 0.14in} c@{\hskip 0.14in} c @{\hskip 0.14in}  c @{\hskip 0.14in} c @{\hskip 0.14in} c  @{\hskip 0.14in} c @{\hskip 0.14in} c @{\hskip 0.14in} c @{\hskip 0.14in} c   }
\hline
 & 1  & 2  & 3  & 4 & 5 & 6  & 7  & 8  & 9  & 10 & 11 & 12 & 13 & 14 & 15 \\
\hline
Quotes &  2.2 &	 3.0 	& 3.0 &	 3.0 	& 2.2 &	 1.0 &	 157.5 &	 242.6 &	 243.8 &	 308.8 	& 287.1 	& 286.2 &	 317.3 	& 264.2 	& 274.8 \\
Trades & 0.3  &	 0.3  &	 0.1 	 & 0.1  &	 0.0  &  	 0.0  &  	 2.0  &	 1.8  &	 2.2  &	 1.9 	 & 2.3 	 & 1.8 &	 1.6  &	 1.5 	 & 1.8  \\
Volume & 23.7 &	 17.2 	& 4.7 &	 2.1 	& 0.2 	 & 0.0 &   	 37.3 	& 34.5 	& 48.7 	& 28.7 	& 24.5 	& 19.5 	& 22.1 	& 10.9 	& 11.7  \\
\hline
\end{tabular}
\end{table}

\section{Data and modelling choices}

\subsection{Data commentary}

\noindent Seeking to model high-frequency Eurodollar futures prices, and hence a component of an interest rate curve, we have collected Thomson Reuters data for continuous contract EDc1 to EDc15 for each US trading day of 2018.  This data reflects Level 1 limit order book best bid-ask-volume quotes, with some additional trade execution data.  The first 15 contracts are a subset of the 44 available market traded contracts through CME group.  In terms of the number of quotes and trades, and total volume traded, we observe greater trading activity in contracts EDc7 to EDc15, relative to the earlier contracts; see Table 1 for reference.  Owing to this market activity, we focus our analysis on  EDc7 to EDc15.

\vspace{0.1 \baselineskip}

\noindent A key step of our data preprocessing is to construct a time-series of microprices for each contract $c$, $P^c := \{ P_t^c  \text{ : }  t \in T^c \}$, such that
$$
P_t^c := \frac{\text{ask volume}_t^c \cdot \text{bid price}_t^c + \text{ask price}_t^c \cdot \text{bid volume}_t^c}{\text{ask volume}_t^c + \text{bid volume}_t^c},
$$

\noindent  following the definition of microprice in \cite{Gath10}.  Table 2 displays various percentile levels for the median daily difference in the high versus the low in $P^c$, offering some intuition around daily variability by contract.  This stands in contrast to the median daily bid-ask spread of 0.5 basis points (bps) for each contract, throughout 2018.

\vspace{0.25 \baselineskip}

\noindent  A second key step is to align and down-sample $\{ P^c : c \in [ \text{EDc7}, \dots, \text{EDc15} ] \}$.  Aligning over index sets $T^c$ is essential given the irregular temporal sampling across contract microprices, while down-sampling conveniently filters for sufficiently small microprice fluctuations, significantly reducing the size of the aligned dataset.  We define a cutoff parameter $M$ and set it to 0.1bps, 
and for each trading day construct a time-series of curve observations $C := \{ \boldsymbol{P}_t = (P_t^c : c \in [ \text{EDc7}, \dots, \text{EDc15} ]) : t \in T \}$ for a common temporal index set $T$, according to the following rules:
\begin{enumerate}
	\item For a new trading day, record an initial curve observation at the first time that there exists a microprice for each contract;
	\item Given an observation time $t^*$, find the earliest time $t>t^*$ such that $| P_{t}^c - P_{t^*}^c| \ge M$ for at least one contract $c$, and record an observation of each contract's most recent microprice at or equal to $t$.  If such a time does not exist, revert to 1. for the next trading day.
\end{enumerate}

%\noindent Finally, we construct a time-series $D :=  \{ \boldsymbol{D}_t = (\boldsymbol{P}_{t+k} : k \in \{ 0, \dots, 99\} ) : t \in T \}$ -- a rolling window of 100 sequential curve observations\footnote{The value `100' is chosen for comparison with \cite{Zhang1}; we leave further refinement as future work.} -- and predict the curve observation at the next event time within the time-series $Y := \{\boldsymbol{Y}_t = \boldsymbol{P}_{t+100} : t \in T \}$.  Further, for each member of $D$, we implement a window normalization whereby each microprice subseries is shifted by its empirical mean within the window, and scaled by the standard deviation of the set of all (shifted) window values.  We normalize the corresponding members of $Y$ by these same statistics.

\noindent Finally, we construct a time-series $D :=  \{ (\boldsymbol{D}_t,\boldsymbol{Y}_t) = \big( (\boldsymbol{P}_{t+k} : k \in \{ -99, \dots, 0\} ), \boldsymbol{P}_{t+1} \big) : t \in T \}$ consisting jointly of a rolling window of 100 sequential curve observations\footnote{The value `100' is chosen for comparison with \cite{Zhang1}, and has been seen to work well in practice.  Note no extensive optimization of this parameter has been performed here.}, and the curve observation at the next event time.  Further, for each member of $D$, we implement a window normalization whereby each microprice subseries in $\boldsymbol{D}_t$ is shifted by its empirical mean within the window, and scaled by the standard deviation of the set of all (shifted) window values.  Similarly, we then normalize the corresponding members of $\boldsymbol{Y}_t$ using these same statistics (computed over the backward looking window).  To be clear, we comment that all no step of our data preprocessing introduces look-ahead bias.

\begin{table}
\centering
\caption{Summary statistics for the yield data used in these notes:  difference in contract microprice levels (bps) high and low on each (US) trading day of 2018;  percentiles and daily standard deviation.  Data source: Thomson Reuters.}
\small\addtolength{\tabcolsep}{-4pt}
\begin{tabular}[c]{c c@{\hskip 0.2in} c @{\hskip 0.2in}  c @{\hskip 0.2in} c @{\hskip 0.2in} c  @{\hskip 0.2in} c @{\hskip 0.2in} c @{\hskip 0.2in} c @{\hskip 0.2in} c   }
\hline
  & EDc7  & EDc8  & EDc9  & EDc10 & EDc11 & EDc12 & EDc13 & EDc14 & EDc15 \\
\hline
min & 0.3   & 1.2 &    1.4     &   1.6  & 1.7 & 1.8 &  1.3 & 1.1 & 1.6  \\
25th perc. &  2.3  & 2.8 &   3.1     &  3.4  & 3.6 & 3.8 &  4.0 & 4.0 & 4.0  \\
50th perc.& 3.0   & 3.7  &  4.3   &  4.6  & 4.8 & 5.0 & 5.1 & 5.1 & 5.3  \\
75th perc. & 4.4   &  5.3  &  5.9   &  6.5  & 6.9 & 6.9 &  7.2 & 7.3 & 7.5 \\
max & 16.2   & 17.1 &   20.6     &   22.1  & 23.3 & 24.9 &  25.8 & 26.2 & 26.2  \\
$\sigma$(daily Hi-Lo) & 2.4   & 2.6  &  2.8  &  3.0  & 3.2 &  3.2  &  3.3 & 3.4 & 3.3  \\
\hline
\end{tabular}
\end{table}

\subsection{Model specification}

\noindent
Given our dataset construction, we model the prediction target $\boldsymbol{Y}_t$ as a multivariate normal random variable with parameters dependent on our window observations:
$$
\boldsymbol{Y}_t \thicksim \text{MVN}( \boldsymbol{\mu}(\boldsymbol{D}_t), \boldsymbol{\Sigma}(\boldsymbol{D}_t)).
$$

\noindent For brevity, we continue by writing $\boldsymbol{\mu}_t := \boldsymbol{\mu}(\boldsymbol{D}_t)$ to denote the mean vector of dimension $1 \times c$ and $\boldsymbol{\Sigma}_t^A := \boldsymbol{\Sigma}(\boldsymbol{D}_t)$ to denote the variance-covariance matrix of dimension $c \times c$ (and with a superscript $A$ to denote ``aleatoric" uncertainty \cite{Ken17}).

\vspace{0.25 \baselineskip}

\noindent We consider two model architectures for estimating $\boldsymbol{\mu}_t$ and $\boldsymbol{\Sigma}_t^A $.  With respect to $\boldsymbol{\mu}_t$, we are inspired to adapt the CNN-LSTM Inc of \cite{Zhang1}, and we seek to compare trading strategy performance to a more straightforward 2-hidden layer multi-layer perceptron (MLP).  Further details for the architectures are offered in Appendix A.3, and a schematic of the model architectures are as depicted in Figure \ref{fg:B}.  A key point of interest, and novel for financial applications, is the network branching, whereby we estimate our targets over two output heads.  We recall \cite{Nix94} that appends an auxillary output to a network architecture to yield a heteroskedatic variance estimate in conjunction with predicting the mean of a univariate Gaussian.  This was extended recently for the multivariate case \cite{Dor18}.  Requiring that $\boldsymbol{\Sigma}_t^A$ is a symmetric positive definite matrix, its inverse is also symmetric positive definite and can be expressed as its Cholesky decomposition.
Hence, writing $(\boldsymbol{\Sigma}_t^A)^{-1} = \boldsymbol{L}_t\boldsymbol{L}_t^T$ for a lower triangular matrix $\boldsymbol{L}_t$, as per \cite{Dor18}, and with minor notional updates for our specific problem, the network loss function $\mathcal{L}$ has been shown to be
\begin{equation}
\mathcal{L} = -2 \Big[ \sum_{i=1}^c \log l_t^{ii} \Big] + (\boldsymbol{Y}_t - \boldsymbol{\mu}_t)^T \boldsymbol{L}_t \boldsymbol{L}_t^T (\boldsymbol{Y}_t-\boldsymbol{\mu}_t).
\end{equation}

\begin{center}
\makebox[\textwidth]{
  \includegraphics[height=3.3in, width=6.6in]{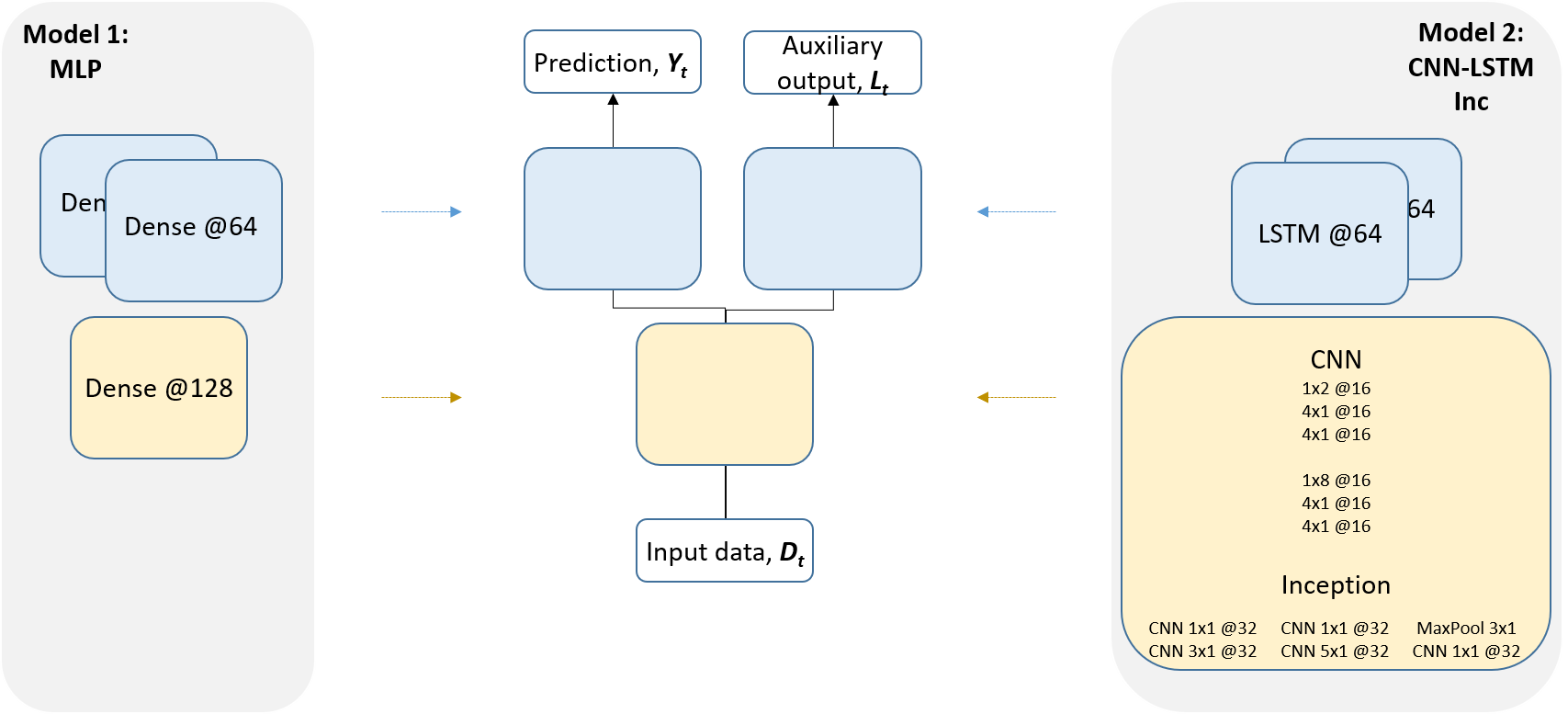}}\par
  \captionof{figure}{Model architectures for predicting curve moves with simultaneous aleatoric uncertainty learning.  The MLP specification was chosen such that it had a similar number of parameters to the CNN-LSTM Inc -- the latter model has nearly $100,000$ parameters, while the former has about $133,000$. Anecdotally, we determined that the branching architecture, featuring a common module before branching (shaded yellow above), improved fit significantly relative to the exclusion of a common module.  \label{fg:B}}
\end{center}

\noindent Here $l_t^{ii}$ denotes the $i$th diagonal entry of $\boldsymbol{L}_t$.  It is interesting to note that $\boldsymbol{\Sigma}_t^A$ is implicitly learnt through the loss function, without reference to explicitly labelled variance-covariance data.  We estimate $\boldsymbol{L}_t$  as output to each model, applying a linear activation function after the final layer.  The diagonal elements are then exponentiated, ensuring the positive definiteness and hence invertibility of $(\boldsymbol{\Sigma}_t^A)^{-1}$, so that $\boldsymbol{\Sigma}_t^A$ is recoverable.

\vspace{0.25 \baselineskip}

\noindent  There are various proposals for approximating epistemic uncertainty for deep neural network models, including pseudo-Bayesian approximations.  One popular technique applies to models estimated with dropout.  

\noindent Recall that dropout was first presented in \cite{Hin12} as a regularization method in an attempt to prevent neural network overfitting and improve model performance on out-of-sample data.  The intuition for implementing dropout is to randomly omit some proportion of hidden weights (by setting them to `0') for each training iteration of a network, and rescaling the final model weights by this proportion at test time.  This was originally interpreted as giving a similar effect to calculating an ensemble of models (depending on which parameters were omitted) and taking their average result as a better predictor than any one model on out-of-sample data.  In the context of modern deep learning, dropout is popular amongst some practitioners for the perceived regularization benefit and for its relative ease of implementation.  The key to retrieving the epistemic uncertainty approximation is the technique of `dropout sampling' \cite{Gal16}.  This technique is applied by generating many model predictions for a single input datum with random dropout sampling enabled at test time.  It has been shown that one can naturally retrieve an approximation to epistemic uncertainty for predictions, calculating the estimate in a straightforward manner by averaging \cite{Gal16}. %; for our model see the formula below.

\vspace{0.25 \baselineskip}

\noindent Hence, we have that for $N$ stochastic dropout samples associated with a single model prediction, our parameter estimates are calculated as:
$$
\widehat{\E \boldsymbol{Y}_t} := \hat{\boldsymbol{\mu}}_t = \frac{1}{N} \sum_{n=1}^N \hat{\boldsymbol{\mu}}_{t,n}  \qquad \text{and } \qquad  \hat{\boldsymbol{\Sigma}}_t^A = \frac{1}{N} \sum_{n=1}^N \hat{\boldsymbol{\Sigma}}_{t,n}^A , \qquad \text{ }
$$

\begin{table}[ht]
\centering
\begin{minipage}{.5\textwidth}
  \centering
%  \begin{tabular}[c]{c @{\hskip 0.2in} c @{\hskip 0.2in}  c @{\hskip 0.2in} c @{\hskip 0.2in}  c}
\resizebox{\textwidth}{!}{%
\begin{tabular}{*5c}

 &  \multicolumn{2}{c}{Loss} & \multicolumn{2}{c}{Mean (Std dev.) SE}\\
  month & MLP & CLI  & MLP & CLI \\
  \cmidrule(lr){1-1} \cmidrule(lr){2-3}  \cmidrule(lr){4-5}

  7 & -4.91 & \bf{-6.74} & \textbf{0.483*} (2.335) & 0.491 (2.275) \\
  8 & -1.83 & \bf{-5.64} & 0.744 (2.691) & \textbf{0.719*} (2.482)  \\
  9 & -1.83 & \bf{-4.93} & \textbf{0.738*} (2.593) & 0.756 (2.474)  \\
  10 & -1.22  & \bf{-4.33} & 0.647 (2.141) & \textbf{0.615*} (2.012)  \\
  11 &  -4.08 & \bf{-6.66} & 0.519 (2.093) & \textbf{0.474*} (2.011)  \\
  12 & - & -  & - & - \\
  \cmidrule(lr){1-1} \cmidrule(lr){2-3}  \cmidrule(lr){4-5}
  Avg &  -2.77  &  \bf{-5.66}  & 0.626 (2.371) & \textbf{0.611} (2.251) \\ 
  \cmidrule(lr){1-1} \cmidrule(lr){2-3}  \cmidrule(lr){4-5}
 & & & \multicolumn{2}{c}{  \begin{tabular}{@{}c@{}} \footnotesize{*p $\le 0.0001$ for each one-tailed}  \\ \footnotesize{test of least model MSE by month} \end{tabular}}  \\
  \end{tabular}}

%  7 & -4.91 & \bf{-6.74} & \bf{0.483} & 0.491 \\
%   &  &  & (\bf{0.483}) & (0.491) \\
%  8 & -1.83 & \bf{-5.64} & 0.744 & \bf{0.719}  \\
%   &  &  & (\bf{0.483}) & (0.491) \\
%  9 & -1.83 & \bf{-4.93} & \bf{0.738} & 0.756 \\
%   &  &  & (\bf{0.483}) & (0.491) \\
%  10 & -1.22  & \bf{-4.33} & 0.647 & \bf{0.615} \\
%   &  &  & (\bf{0.483}) & (0.491) \\
%  11 &  -4.08 & \bf{-6.66} & 0.519 & \bf{0.474} \\
%   &  &  & (\bf{0.483}) & (0.491) \\
%  12 & - & -  & - & - \\
%   &  &  &  &  \\

\end{minipage}%
\begin{minipage}{.5\textwidth}
  \centering
  \includegraphics[width=1.05\linewidth]{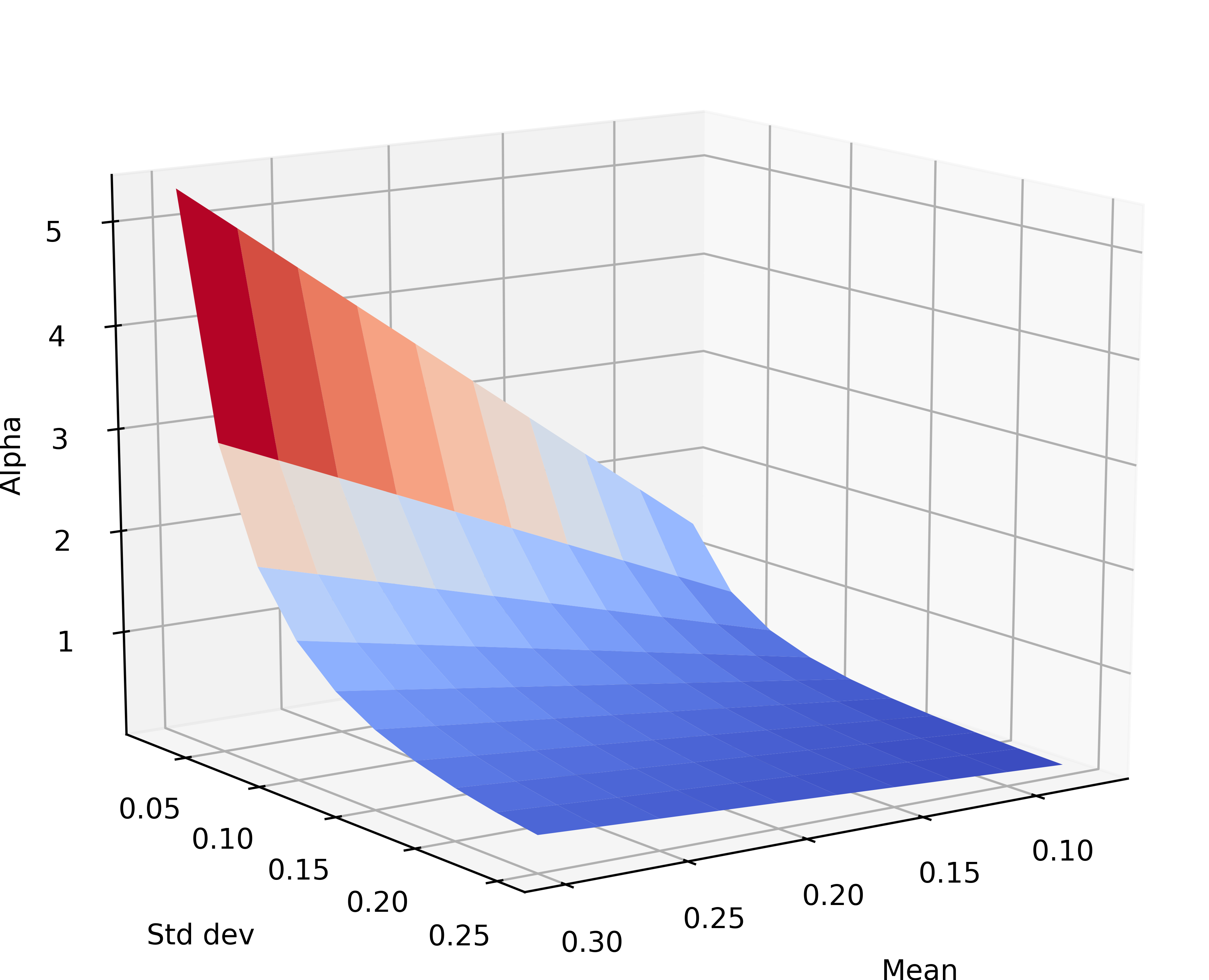}
  
\end{minipage}
\captionof{figure}{Left: Validation loss and standard error statistics by month for MLP and CNN-LSTM Inc models.  Each model is estimated using the full year's data, up to but excluding the validation month. Right:  $\alpha$-surface of Section 2.2, with $\alpha$ set to 1 for $(\mu,\sigma) = (0.3,0.1)$.}
\label{fg:C}
\end{table}

\noindent with the expectation estimate according to \cite{Gal16}, and where the covariance estimate conforms to \cite{Rus19}.  Again, as in \cite{Rus19}, an estimate of the epistemic uncertainty of the prediction is given by
$$
\hat{\boldsymbol{\Sigma}}_t^E = \frac{N}{N-1} \cdot \Bigg( \frac{1}{N}  \sum_{n=1}^N \hat{\boldsymbol{\mu}}_{t,n} \hat{\boldsymbol{\mu}}_{t,n}^T - \hat{\boldsymbol{\mu}}_t \hat{\boldsymbol{\mu}}_t^T \Bigg),
$$
where we have made an additional adjustment to yield an unbiased sample covariance estimate.  Hence the total prediction uncertainty estimate can be written
$$
\widehat{\text{Var}(\boldsymbol{Y}_t)} := \hat{\boldsymbol{\Sigma}}_t  = \hat{\boldsymbol{\Sigma}}_t^A + \hat{\boldsymbol{\Sigma}}_t^E.
$$

\subsection{Trading strategy} 

\noindent In this section, we present a trading strategy that depends on our model outputs $\hat{\boldsymbol{\mu}}_t$ and $\hat{\boldsymbol{\Sigma}}_t$ for selecting $\alpha_t$, the investment size for a particular trade at each decision time $t$.

% Our model outputs, $\hat{\boldsymbol{\mu}}_t$ and $\hat{\boldsymbol{\Sigma}}_t$, can effect a backtest through $\alpha$, our investment size for a particular trade.

\vspace{0.25 \baselineskip}

\noindent  In the literature, a typical backtest given a classification or regression model often amounts to selecting a trading decision based on one of 3 choices: sell 1 unit, do nothing or buy 1 unit, respectively corresponding to $ \alpha \in \{ -1,0,1\} $.  However, a more flexible range of permissible investment sizes, such as the case of continuous $\alpha \in [-1,1]$, could lead to trading strategies showing non-trivial improvement across real-world performance outcomes.  When interpreting $\hat{\boldsymbol{\mu}}_t$ and $\hat{\boldsymbol{\Sigma}}_t$ as true parameters of the underlying return distribution, one can intelligently scale relative investment sizing depending on a pre-defined trading objective.  An example objective that we present is to maximise the out-of-sample Sharpe ratio trading performance metric.  Assuming additive rather than multiplicative returns, and further assuming the simplified setting whereby trade opportunities present for each estimated pair of distribution parameters a uniformly equal number of times, we show in Appendix A.1 that the optimal investment size is given by
\[
\alpha_{t,i} = \frac{ \hat{\mu}_{t,i}} { \hat{\sigma}_{t,i}^2}.
\]  
Here, $\mu_{t,i}$ denotes the $i$th element of $\hat{\boldsymbol{\mu}}_t$, and $\hat{\sigma}_{t,i}$ denotes the square root of the $i$th diagonal element of $\hat{\boldsymbol{\Sigma}}_t$.

\begin{table}[h]
\centering
\caption{Monthly Sharpe performance for MLP and CNN-LSTM Inc models for a set of investment sizing strategies, on out-of-sample data.  Cumulative 5 month Sharpe is shown in the last row of each table.
Investment size is scaled by realised window volatility (\textit{Rlsd vol}), the aleatoric uncertainty estimate (\textit{Alea}), and the sum of aleatoric and epistemic uncertainty (\textit{Al$+$Ep}).  \textit{Base} has no investment sizing by uncertainty.}
\small\addtolength{\tabcolsep}{-4pt}
\begin{tabular}[c]{c @{\hskip 0.2in} c @{\hskip 0.2in}  c@{\hskip 0.2in} c@{\hskip 0.2in} c @{\hskip 0.2in} c  @{\hskip 0.2in} c  @{\hskip 0.2in}  c   @{\hskip 0.2in}  c @{\hskip 0.2in}  c @{\hskip 0.2in}  c}
   &    & \multicolumn{4}{c}{MLP} & \multicolumn{4}{c}{ CNN-LSTM Inc} \\

   &    & \multicolumn{4}{c}{} & \multicolumn{4}{c}{ } \\
&  & Rlsd vol & Base  & Alea  &  Al$+$Ep &  Rlsd vol & Base  & Alea  &  Al$+$Ep \\
\cmidrule(lr){3-6}\cmidrule(lr){7-10}
&7 & - & - & - &    -     & -  & - & - & - \\
Diagonal&8 & 0.87 & 0.81 & 0.88 &    \bf{0.89}      &  \bf{1.01} &  0.88 &  0.97 & 0.98  \\
$\boldsymbol{\Sigma}$ &9 & 1.99 & \bf{2.32}  & 1.98  &  2.00   &   1.39  &  1.64  &  \bf{1.86}  & 1.81 \\
&10 & 1.76  & \bf{2.47} &  1.96 &   1.98   &  1.75 &  \bf{2.45} & 2.25 &  2.23  \\
&11 & \bf{1.33}  & 1.14 & 1.21  &    1.21   & \bf{1.31} & 1.20 & 1.23 & 1.25 \\
&12 & 2.69 & 2.90  & 2.98  &   \bf{2.99}   &  2.57 &  \bf{2.99} &  2.95 &  2.94 \\
\cmidrule(lr){3-6}\cmidrule(lr){7-10}
&Cuml &  1.29  &  1.40  &  1.42  &  \bf{1.44}  &   1.24 &  1.27 &  1.37 & \bf{1.38} \\
\cmidrule(lr){3-6}\cmidrule(lr){7-10}

   &    & \multicolumn{4}{c}{} & \multicolumn{4}{c}{ } \\

\cmidrule(lr){3-6}\cmidrule(lr){7-10}
& 7 & - & - & - &    -     & -  & - & - & - \\
Full  & 8 & 0.89 & 0.83 & 0.95 &    \bf{0.96}      &  0.95 &  0.82 &  1.01 & \bf{1.02}  \\
$\boldsymbol{\Sigma}$& 9 & 1.84 & \bf{2.19}  & 2.12  &  2.12   &   1.65  &  1.85  &  \bf{2.17}  &  2.15 \\
& 10 & 1.71  & \bf{2.47} &  1.84 &   1.88   &  1.73 &  \bf{2.46} & 2.45 &  2.41  \\
& 11 & \bf{1.34}  & 1.12 & 1.24  &    1.24   & \bf{1.27} & 1.15 & 1.18 & 1.20\\
& 12 & 2.65 & 2.87  & 2.87  &   \bf{2.87}   &  2.57 &  \bf{2.94} &  2.91 &  2.91 \\
\cmidrule(lr){3-6}\cmidrule(lr){7-10}
& Cuml &  1.25  &  1.38  &  1.39  &  \bf{1.42}  &   1.27  &  1.26 &  \bf{1.42} & 1.41 \\
\cmidrule(lr){3-6}\cmidrule(lr){7-10}
\end{tabular}
\end{table}

\vspace{0.25 \baselineskip}

\noindent To assist intution, we chart the corresponding $\alpha$-surface on the right-hand panel of Figure \ref{fg:C}, where we have made the additional arbitrary choice to rescale the surface such that the point $(\mu, \sigma)= (0.3,0.1)$ corresponds to $\alpha =1$.

\vspace{0.25 \baselineskip}

\noindent  There are two pleasing outcomes of this approach relative to \cite{Zhang2}, an early work to utilize an approximation to Bayesian uncertainty estimation for trading given the predictions of a deep classification model.  Firstly, we do not increasingly penalise a trade based on increased uncertainty alone, but rather based on risk-adjusted returns.  This corresponds to real-life, where high uncertainty for a trade is not necessarily outright undesirable -- a trade proposition should have its estimated uncertainty considered in conjunction with the expected return size.  Secondly, we do not depend on an ad hoc rules based approach for setting $\alpha_t$, and instead select it in an arguably more intuitively pleasing way.

\section{Results}

Given we have one full year of data, we partition by month and evaluate and verify our models by training on months $1$ to $M-1$, validating using month $M$, and finally testing our trading strategy using month $M+1$, for $M \in \{7,8,9,10,11\}$.  This designation is arbitrary, but somewhat logical within, say, a trading platform that may be updated on a monthly basis.

\vspace{0.25 \baselineskip}

\noindent  All models are estimated using Keras \cite{chollet2015keras}.  We trained with the objective of minimising validation loss, and implemented an early stopping rule with a patience of 15, restoring the best model weights.  Further modelling details are given in Appendix A.3.  Of important note, we trained the MLP and CNN-LSTM Inc with diagonal variance-covariance matrices to compare to the proposed `full' variance-covariance case.  Also, for practicality, we tuned an L2 weight regularisation parameter, and selected a uniform dropout rate,

\begin{center}
\makebox[\textwidth]{
  \includegraphics[height=3.1in, width=6.6in]{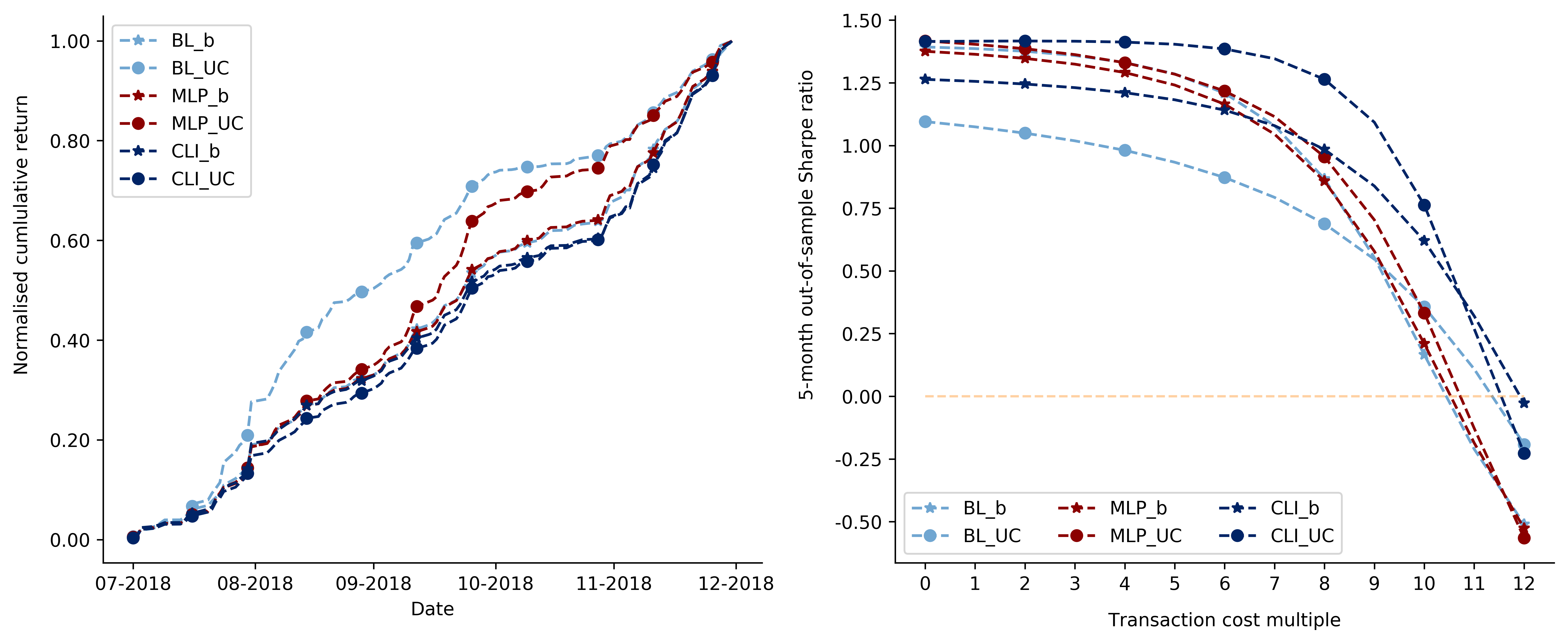}}\par
  \captionof{figure}{Left: Out-of-sample normalised daily cumulative basis point return for \textit{Base} and \textit{Al$+$Ep} long/short trading strategies for the baseline, MLP and CNN-LSTM Inc models.  Right: 5-month out-of-sample Sharpe ratio for \textit{Base} and \textit{Al$+$Ep} long/short trading strategies for the baseline, MLP and CNN-LSTM Inc models, accounting for transaction costs.  The $x$-axis varies by multiple of transaction costs, where a multiple of `1' represents a cost of $\frac{1}{20}^{\text{th}}$ of the trading threshold of $0.1$bps.
\label{fg:D}}
\end{center}

\noindent by grid search over the validation data, for the MLP with diagonal covariance model.  Hence, these parameter values were respectively set to 1e-8 and 0.1 for each of the four models\footnote{\noindent Where applicable, dropout layers are included as per \cite{Gal16}; for convolutional and recurrent layers we follow \cite{GalCNNDO} and \cite{GalRNNDO}, with the exception that we add no dropout within the inception layer.  We calculate 30 dropout samples for each prediction.  Further, we found that trading strategy performance showed negligible difference using 60 samples.}.  Other important details of a linear Bayesian baseline model, that we offer as a relevant benchmark when discussing out-of-sample trading performance, are included in Appendix A.3.

\vspace{0.25 \baselineskip}

\noindent  In the left-hand table of Figure \ref{fg:C}, we see that the validation data loss is uniformly lower for CNN-LSTM Inc relative to the MLP model, on average differing by a factor of $2.0$.  On a mean-square error basis, the CNN-LSTM Inc outperforms the MLP for 3 out of 5 months for a one-tailed $t$-test of least MSE, with p-values $\le 0.0001$ in each case.  On average across all months, the CNN-LSTM Inc MSE is 2.5\% smaller.  The outperformance of the CNN-LSTM Inc model in these respects supports the claim of the usefulness of this model for the prediction task.

\vspace{0.25 \baselineskip}

\noindent Next, we discuss the results for the trading strategy described in section 2.3, focussing on the monthly and cumulative Sharpe performance on out-of-sample data.  Firstly, for all strategies, we note that we have defined a trading threshold of 0.1bps such that any trade entered must have a predicted (absolute) asset price change greater than or equal to the threshold.  This threshold was chosen to be equal to the cutoff parameter defined in section 2.1.  Next, we define a strategy called \textit{Base} that enters a trade position $\alpha \in \{-1,1\}$ depending on the sign of the predicted price change (respectively, lower or higher, and also corresponding to short or long).  We also analyse a trading strategy that takes the notion of prediction uncertainty into account as proposed in section 2.3, and explore 2 additional substrategies whereby we estimate uncertainty by alternative methods.  For the strategy \textit{Rlsd vol}, the realised volatility of the input data window is used as a proxy for uncertainty, and serves as an alternative benchmark.  For \textit{Alea}, aleatoric uncertainty is instead used, and for \textit{Al$+$Ep}, we use the sum total of aleatoric and epistemic uncertainty as per section 2.2.  We do

\begin{center}
\makebox[\textwidth]{
  \includegraphics[height=3.1in, width=6.6in]{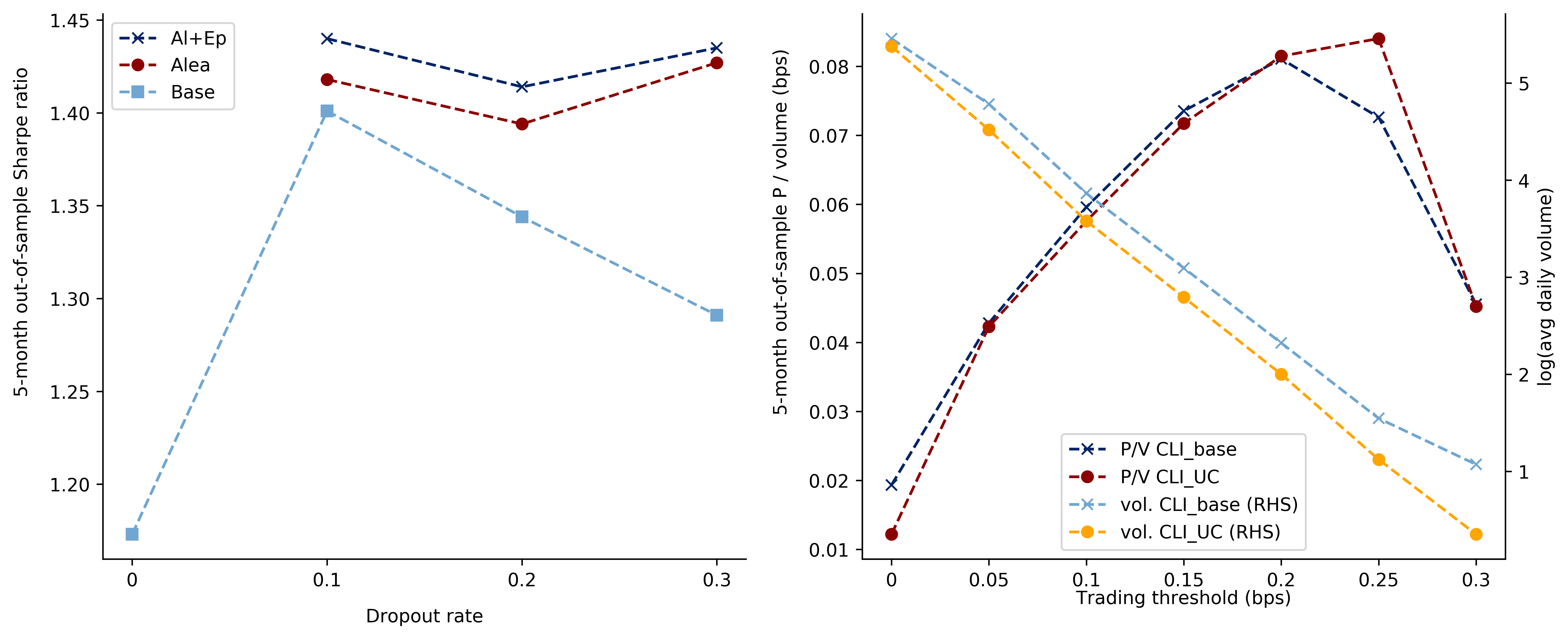}}\par
  \captionof{figure}{Left: 5-month out-of-sample Sharpe ratio for trading strategy based on long/short positions without uncertainty (\textit{Base}), with aleatoric uncertainty only (\textit{Alea}), and using our full uncertainty estimate (\textit{Al$+$Ep}), for varying dropout rates, given a diagonal covariance MLP model.  Right: 5-month out-of-sample profit over volume (left-hand axis) for varying trading thresholds ($x$-axis) for the full covariance CNN-LSTM Inc model.  The right-hand axis the log average daily volume traded for each trading threshold.
\label{fg:E}}
\end{center}

\noindent not present results for an analogous strategy based on epistemic uncertainty alone, on finding cases of unreasonably large $\alpha$ corresponding to an uncertainty estimate for a particular prediction very close to 0.

\vspace{0.25 \baselineskip}

\noindent The out-of-sample Sharpe ratios are shown for the performance of each model by investment sizing strategy in Table 3.  For comparison, we show the results for both the diagonal (top row) and full (bottom row) variance-covariance model variants.  For three of the four subtables, we notice that the cumulative 5-month Sharpe performance is greatest for the \textit{Al$+$Ep} strategy, with \textit{Al} second greatest in each case (with outperformance between 0.7 and 2.2$\%$).  For the fourth subtable, \textit{Al} is greatest and marginally ahead of \textit{Al$+$Ep} (by 0.7$\%$).  More strikingly, the relative outperformance across \textit{Al$+$Ep} strategies compared to \textit{Base} ranges from 2.9 to 12.7$\%$, and we observe a minimum 11.3$\%$ outperformance relative to \textit{Rlsd vol)}.  These results offer a preliminary demonstration of the utility of incorporating estimates of both aleatoric and epistemic uncertainty in the context of investment sizing within our high-frequency trading setting.

\vspace{0.25 \baselineskip}

\noindent Of interest, we highlight what on one hand appears to be the outperformance of the diagonal MLP \textit{Al$+$Ep} strategy versus any other, and on the other hand the apparent strong performance of the Bayesian OLS \textit{Base} strategy as shown in the Appendix in Figure \ref{fg:F}.  To rationalise this, we first consider the sample paths of returns as in the left-hand panel of Figure \ref{fg:D}.  We observe that the out-of-sample daily cumulative (basis point) return of the presented strategies appear mostly without outliers and exhibit relatively smoothness, and we do not attribute any outperformance due to any obvious lucky or outsized earnings.  Instead, we next consider the stylised impact of transaction costs for comparing model and strategy outperformance.

\vspace{0.25 \baselineskip}

\noindent Firstly, we define an element of total net transaction cost to be equal to 0.005bps per unit volume.  Based on our trading threshold of 0.1bps, this corresponds to 5\% of the minimum prediction size such that $\vert \alpha_t \vert>0$ for any trade.  We acrue the transaction cost in proportion to volume transacted at each time step.  To be clear, we only trade the delta-adjustment to achieve the required position of size $\alpha_t$ for the specific asset, relative to the the existing position on the day's most recent trade time.  From the right-hand panel of Figure \ref{fg:D} we see 5-month out-of-sample Sharpe ratio for \textit{Base} and \textit{Al$+$Ep} long/short trading strategies for the linear baseline, MLP and CNN-LSTM Inc models, accounting for multiples from 0 to 12 of our defined element of transaction cost.  To be clear, the label `0' denotes the absence of transaction costs, and the upper label of `12' is the minimal integral multiple of transaction costs such that the Sharpe ratios for each presented strategy is negative.  From this Figure it is immediately clear that the CNN-LSTM Inc model with uncertainty outperforms all other models at most levels of transaction costs, and that the outperformance is greatest at the middle of the range.  This analysis contributes somewhat to understanding the relative utility between alternative models, and strongly vindicates CNN-LSTM Inc with uncertainty for trading.  Of course, the careful reader may notice that the transaction costs analysed here are significantly less than the median daily bid-ask spread of 0.5bps quoted in section 2.1.  With respect to this observation, it is important to understand that the approach presented may not necessarily be traded outright as a stand alone strategy in practice.  The utility of the approach may be derived from its addition to a broader market making or execution strategy, where trading is occuring anyway.  Hence a deeper understanding of the impact of transaction costs, especially relative to the bid-ask spread, may depend on a more realistic and sophisticated trading framework.  We leave these considerations for future work.

\vspace{0.25 \baselineskip}

\noindent Other points of interest around our backtest include observing how the choice of dropout rate or trading threshold impacts out-of-sample performance.  In the case of varying the dropout rate, for the diagonal MLP, we tested cumulative 5-month Sharpe for varying dropout rates for \textit{Base}, \textit{Al$+$Ep} and \textit{Al}, and chart the results in the left-hand panel of Figure \ref{fg:E}.  Interestingly, we see that for either of the two trading strategies utilizing uncertainty, the Sharpe ratios are relatively stable, ranging by less than 1.8$\%$ each across our three tested non-zero dropout rates.  However, similar stability is clearly not evident in the case of the \textit{Base} strategy, which achieves peak Sharpe performance at the choice of dropout rate of 0.1, but displays sharp decrease in performance for increasing dropout rates, and when dropout is not utilized at all.  One may wonder to what extent this trading performance stability for varying dropout rates might be observed in increasingly sophisticated models such as CNN-LSTM Inc.  If so, this could dominate the need for a large degree of tuning of the dropout rate when estimating the model -- this is an increasingly resource consuming task with such model complexity.  We leave this as a consideration for future work.

\vspace{0.25 \baselineskip}

\noindent Next we consider the case of varying the trading threshold, and refer to our results in the right-hand panel of Figure \ref{fg:E}.  Here we see 5-month out-of-sample profit over volume (left-hand axis) for varying trading thresholds ($x$-axis) for the full covariance CNN-LSTM Inc model.  This statistic can be alternatively recognised as the breakeven net transaction cost for a strategy, with higher values corresponding to strategies with greater average net profit per trade.  It is interesting to note that we can optimize the threshold to maximize the Sharpe ratio, though we leave further consideration of this to the interested practioner.  For added interest, we also plot the log average daily volume traded for each trading threshold on the right-hand axis, and see that it is approximately log linearly decreasing with an increasing threshold.

\vspace{0.25 \baselineskip}

\noindent Finally, by the right-hand panel of Figure \ref{fg:F} in the Appendix, we plot estimated model reward-risk ratio versus realized monthly Sharpe on out-of-sample data, for the diagonal MLP.  In that this plot is sufficiently representative of each of the four models, we infer that despite the obvious underperfomance in the tails (where sample sizes are relatively small anyway) the models appear somewhat sensibly calibrated, though we leave any other tuning in this respect as future work.

\section{Conclusion}

We present a model for price change prediction of the Eurodollar Futures curve, as a function of a recent history of asset price data, within a high-frequency domain.  In doing so, we extend the existing state-of-the-art model deep learning architecture of \cite{Zhang1} to multivariate, correlated price data, and improved price prediction relative to benchmark models, for small time horizons and in a regression setting.  Importantly, we develop a preliminary analysis describing the estimation of deep learning model-generated uncertainty for financial prediction, and further showed how these uncertainties can be useful for investment size scaling within trading strategies.  There is future work that could improve on this analysis, with much of it potentially important for applications at an industrial scale.  With respect to data, a larger subset of available Eurodollar futures contracts could be used, as could alternative, correlated asset prices or economic information.  One could also include deeper order book data, such as Level 2 quotes.  With respect to the models used, there is more to be done in optimally engineering model architecture for aleatoric uncertainty estimation.  Alternative methods for estimating epistemic uncertainty could be explored.  On one hand, concrete dropout could be attempted for automating the tuning of the dropout rate \cite{GalCD}.  On the other, Hamiltonian Monte Carlo methods could be explored as an alternative to estimating epistemic uncertainty, and potentially applied, for example, with {\small {\tt hamiltorch}} \cite{Cobb19_2}.  Finally, with respect to the realities of implementing a trading strategy in practice, we allude to three main avenues of further research;  a deeper appreciation of the impact of transaction costs beyond our preliminary analysis, the return to fine-tuning the dropout rate if performance metrics around trading are relatively stable, and finally further consideration of model calibration.

\section*{Acknowledgements}

The authors would like to thank the Oxford-Man Institute of Quantitative Finance for its generous support, including data access.  Trent Spears would further like to thank Dr Jan-Peter Calliess and Prof Nir Vulkan for their helpful insights and guidance, as well as the student members of the OMI, especially Bryan Lim, for their suggestions and encouragement.

\newpage
\bibliographystyle{unsrtnat}
\setlength{\bibsep}{1pt}
\footnotesize\bibliography{bibs}{}

\newpage

\appendix
\small

\section{Appendix}

\subsection{Choosing $\alpha$ -- sketch intuition}
Assume we are given $N$ trading opportunities $\{X_i\}$ with normally distributed and uncorrelated returns, parameterised by means and variances given by $(\mu_i,\sigma_i^2)$ for each choice $i$.  Also assume these parameters take strictly positive values, and that we trade each opportunity an equal number of times.  We seek to trade an optimal amount $\alpha_i$ of $X_i$ to maximises the Sharpe ratio $S$ of the return.  We write
$$
S = \frac{ \sum \alpha_j \mu_j}{\sqrt{\sum \alpha_j^2 \sigma_j^2}}.
$$
Taking the derivative of $S$ with respect to $\alpha_i$ yields
$$
\frac{dS}{d\alpha_i} = \frac{\mu_i \sum_{j \ne i} \alpha_j^2 \sigma_j^2 - \alpha_i \sigma_i^2 \sum_{j \ne i} \alpha_j \mu_j }{(\sum \alpha_j^2 \sigma_j^2)^{3/2}}
$$
which equals 0 for
$$
\alpha_i^* = \frac{\mu_i / \sigma_i^2}{\sum_{j \ne i} \alpha_j^* \mu_j / \sum_{j \ne i} \alpha_j^{2*} \sigma_j^2}.
$$
One can see that a solution for each $i$ is given by
$$
\alpha_i^* = \frac{\mu_i }{\sigma_i^2}.
$$
Taking the second derivative of $S$ with respect to $\alpha_i$ yields
$$
\frac{d^2S}{d\alpha_i^2} = \frac{-\sigma_i^2 (\sum_{j \ne i} \alpha_j \mu_j \sum \alpha_j^2 \sigma_j^2 + 3 \alpha_i \mu_i \sum_{j \ne i} \alpha_j^2 \sigma_j^2 -  3 \alpha_i^2 \sigma_i^2 \sum_{j \ne i} \alpha_j \mu_j)}{(\sum \alpha_j^2 \sigma_j^2)^{5/2}}.
$$
Substituting $\alpha_i^*$ we have that
\begin{align*}
\frac{d^2S}{d\alpha_i^2}\Big\vert_{\alpha_i=\alpha_i^*} 
&= \frac{-\sigma_2^2 \Big(   \sum_{j \ne i} \mu_j^2 / \sigma_j^2 \sum \mu_j^2 / \sigma_j^2  + 3 \mu_i^2 / \sigma_i^2 \sum_{j \ne i} \mu_j^2 / \sigma_j^2 - 3 \mu_i^2 / \sigma_i^2 \sum_{j \ne i} \mu_j^2 / \sigma_j^2 \Big)}{(\sum \mu_j^2 / \sigma_j^2)^{5/2}} \\
&= \frac{-\sigma_2^2 \cdot   \sum_{j \ne i} \mu_j^2 / \sigma_j^2  }{(\sum \mu_j^2 / \sigma_j^2)^{3/2}} \\
& < 0,
\end{align*}
hence for this optimal $\alpha_i^*$ the Sharpe ratio is indeed maximised.

\subsection{Additional data commentary}

\noindent \textit{Futures dataset}.  The set of 44 Eurodollar futures contracts consist of 40 quarterly contracts and 4 serial contracts;  the former are based on a quarterly cycle (beginning in March), while the serial contracts are for the first four closest months (with respect to roll dates) outside of that cycle.  Note that by this definition, the serial contracts must be contained in EDc1 to EDc6.  
Hence, depending on the month of the year, our focus on EDc7 to EDc15  represents a segment of an interest rate curve starting 7, 8 or 9 months in the future, as well as the following 2 years. 

\vspace{0.1 \baselineskip}

%We note that the microprices are constructed after removing null data and 3 `bad' days  [These 3 bad days are such that raw data for EDc7 is null.].  
\noindent  \textit{Microprice}.  There are well justified reasons to model using the microprice -- including for its utility in excess of using the simpler midprice for capturing an element of volume imbalance between each side of the order book.  In our case, it also stands to consider microprices from a practical point of view given data quality.

\vspace{0.1 \baselineskip}

% check QB > Almgren > white papers (SZ 24/6). Oct19 also have the JPM paper.

\vspace{0.1 \baselineskip}

\noindent \textit{Dataset size}.  We collected 25GB of raw data as .csv files for our target contracts, and have prepared just over 50GB in .h5 files on conclusion of preprocessing (and accounting for windows of input data).

\subsection{Modelling notes}

\textit{Initial exploration}.  We compare a suite of models for predictive performance on out-of-sample data to develop intuition about suitable rates curve model architectures.  This is independent of considering any uncertainty measures.  Listing these models by increasing complexity, we explore simple and exponentially-weighted averaging, ordinary least squares, principal component analysis, single- and multi-layer perceptrons, CNN, LSTM, CNN-LSTM, and finally a CNN-LSTM Inc in the spirit of \cite{Zhang1}.  We evaluate model fit by comparing mean-squared error and Huber loss \cite{Hub64} on our validation data, whereby the proceeding observations were broadly consistent.  The first clear result is that for the $t+1$ prediction horizon, there is a clear trend in significantly decreasing prediction error as model complexity increases.  Further, the CNN-LSTM Inc always outperforms every other model for both error types, which complements the results of \cite{Zhang1}.  Secondly, each model broadly displays larger error for larger prediction horizons, which accords with intuition.  Considering the $t+100$ prediction horizon, there is such little variation in error across rows that it is clear that not only do we we have a poor ability to forecast price action relatively far into the future, but all models appear to be of about as little utility as each other.  An inflection point for these observations seems apparent at the $t+10$ prediction horizon.

\vspace{0.25 \baselineskip}

\noindent \textit{CNN-LSTM Inc model details}.  The model architecture is as shown in Figure \ref{fg:B}, and we follow our reference closely, since we have no strict advantage in selecting this beyond intution.  A key difference for our setting, however, is that we do not include any information from the order book beyond Level 1, and the spatial component analogue for our model is equal to adjacent asset prices on the curve, rather then levels of limit order book data on the single asset.  Another difference is that our reference work's initial 3 convolutional layers, that convolve price and volume levels at different depths of the limit order book, are not relevant in our case, and so are excluded.  To recapitulate some of the other details, we use 6 convolutional layers of 16 filters each and with kernel sizes, respectively, (1,2), (4,1), (4,1), (1,8), (4,1), (4,1).  All layers have a leaky ReLU output with parameter $\alpha = 0.01$.  Same padding is applied to the latter two layers.  The inception layer is a 2x3 tower whereby the input layer consists of a MaxPooling layer with pool size (3,1) and stride of 1, and 2 convolutional layers with kernel size (1,1) and 32 filters.  The output layer of the inception network is 3 more convolutional layers with kernel sizes (1,1), (3,1) and (5,1) respectively, all with 32 filters.  All inception modules have same padding, and the convolutional layers have ReLU activation.  The LSTM layers have 64 units, with a subsequent dense layer, and linear output activation function.  Finally, for replicability sake we add that the MLP model uses ReLU activations at all dense layers but the last, where a linear activation function is instead implemented.

\vspace{0.1 \baselineskip}

\noindent \textit{Bayesian linear baseline model details}.  We follow the Bayesian framework for multioutput regression outlined in Appendix A.2 of \cite{Leb12} in estimating a suitable baseline model for comparing our results.  Similarly to section 2.2 of this paper, we model $\boldsymbol{Y}_t \thicksim \text{MVN}( \boldsymbol{\mu}_t, \boldsymbol{\Sigma})$ where $\boldsymbol{\mu}_t = \boldsymbol{D}_t \boldsymbol{\beta}$ is an affine function.  [Hence we have adjusted our earlier definition of $\boldsymbol{D}_t$ by appending a constant term.]  We make the natural assumption that the prior distribution for $\boldsymbol{\Sigma}$ is inverse-Wishart, that is, $\boldsymbol{\Sigma} \thicksim W^{-1}(\boldsymbol{\Omega}, \nu_0)$.  We further assume that the prior for $\boldsymbol{\beta} | \boldsymbol{\Sigma}$ is matrix normal, that is, $\boldsymbol{\beta} | \boldsymbol{\Sigma} \thicksim N_{(100c+1) \times c}(\boldsymbol{\beta_0}, \boldsymbol{\Sigma}, \boldsymbol{\Sigma}_0)$.  Hence, the posterior predictive of $\boldsymbol{Y}_t$ given $n$ historical data observations $\mathcal{D} := (\boldsymbol{D},\boldsymbol{Y}) = \{(\boldsymbol{D}_i,\boldsymbol{Y}_i) : 1 \le i \le n \}$ and a single new observation $\boldsymbol{D}_t$ can be shown to be a multivariate student distribution:
$$
\boldsymbol{Y}_t | \boldsymbol{D}_t, \mathcal{D} \thicksim T((\boldsymbol{\Omega}+\boldsymbol{A^*}).C^{-1}, n+\nu_0),
$$
for matrices $\boldsymbol{A^*} = \boldsymbol{Y}^T\boldsymbol{Y} + \boldsymbol{\beta_0}^T \boldsymbol{\Sigma_0}^{-1} \boldsymbol{\beta_0} - (\boldsymbol{D}^T\boldsymbol{D} \boldsymbol{\hat{\beta}} + \boldsymbol{\Sigma_0}^{-1} \boldsymbol{\beta_0})^T(\boldsymbol{D}^T\boldsymbol{D}+\boldsymbol{\Sigma_0}^{-1})^{-1}(\boldsymbol{D}^T\boldsymbol{D}\boldsymbol{\hat{\beta}}+\boldsymbol{\Sigma_0}^{-1}\boldsymbol{\beta_0})$ and $\boldsymbol{\hat{\beta}} = (\boldsymbol{D}^T\boldsymbol{D})^{-1}\boldsymbol{D}^T\boldsymbol{Y}$, and scalars $C^{-1}=1+ \boldsymbol{D}_t (\boldsymbol{D}^T\boldsymbol{D}+\boldsymbol{\Sigma_0}^{-1})^{-1} \boldsymbol{D}_t^T$, and $\nu_0 = n_0 - (c+(100c+1)) + 1$.  Hence, the required output prediction and uncertainty estimates can be written
\begin{align*}
\widehat{\E[ \boldsymbol{Y}_t | \boldsymbol{D}_t, \mathcal{D} ]} &=  \boldsymbol{D}_t (\boldsymbol{D}^T\boldsymbol{D} + \boldsymbol{\Sigma_0}^{-1})^{-1}(\boldsymbol{D}^T \boldsymbol{Y} + \boldsymbol{\Sigma_0}^{-1} \boldsymbol{\beta_0}), \qquad \text{ and }\\
\widehat{\text{Var}[\boldsymbol{Y}_t | \boldsymbol{D}_t, \mathcal{D}]} &= \frac{1}{n + \nu_0 - 2} (\boldsymbol{\Omega} + \boldsymbol{A^*}) .C^{-1}.
\end{align*}

\vspace{0.1 \baselineskip}

\noindent \textit{Other notes}.  We implement all models in Keras \cite{chollet2015keras}.  We utilize the Adam optimization algorithm when training models \cite{Kin17}.
We batch-train our models
% by calling a custom data generator via the model.fit\_generator function and
whereby each batch contains 1024 samples, with the whole year of data consisting of about 7450 batches.  All results were generated using Tensorflow-GPU 1.9.0 on a 6 core Xeon W-2133 3.2GHz / 12GB NVIDIA GTX 1080 ti / 64GB RAM machine.

\begin{table}[ht]
\centering
\begin{minipage}{.5\textwidth}
  \centering
%  \begin{tabular}[c]{c @{\hskip 0.2in} c @{\hskip 0.2in}  c @{\hskip 0.2in} c @{\hskip 0.2in}  c}
\small\addtolength{\tabcolsep}{-4pt}
\begin{tabular}[c]{c @{\hskip 0.2in} c @{\hskip 0.2in}  c@{\hskip 0.2in} c@{\hskip 0.2in}  c}
   &   & \multicolumn{2}{c}{Bayesian OLS}  \qquad \\

   &    & \multicolumn{2}{c}{} \\
&  & Base & Al$+$Ep \\
\cmidrule(lr){3-4}
&7 & - & -     \\
Full &8 & 0.90 & 0.85   \\
$\boldsymbol{\Sigma}$ &9 & 2.23 & 1.46  \\
&10 & 2.51  & 1.95  \\
&11 & 1.32  & 0.90  \\
&12 & 2.69 & 2.93  \\
\cmidrule(lr){3-4}
&Cuml &  1.39  &  1.10  \\
\cmidrule(lr){3-4}

\end{tabular}

\end{minipage}%
\begin{minipage}{.5\textwidth}
  \centering
  \includegraphics[width=1 \linewidth]{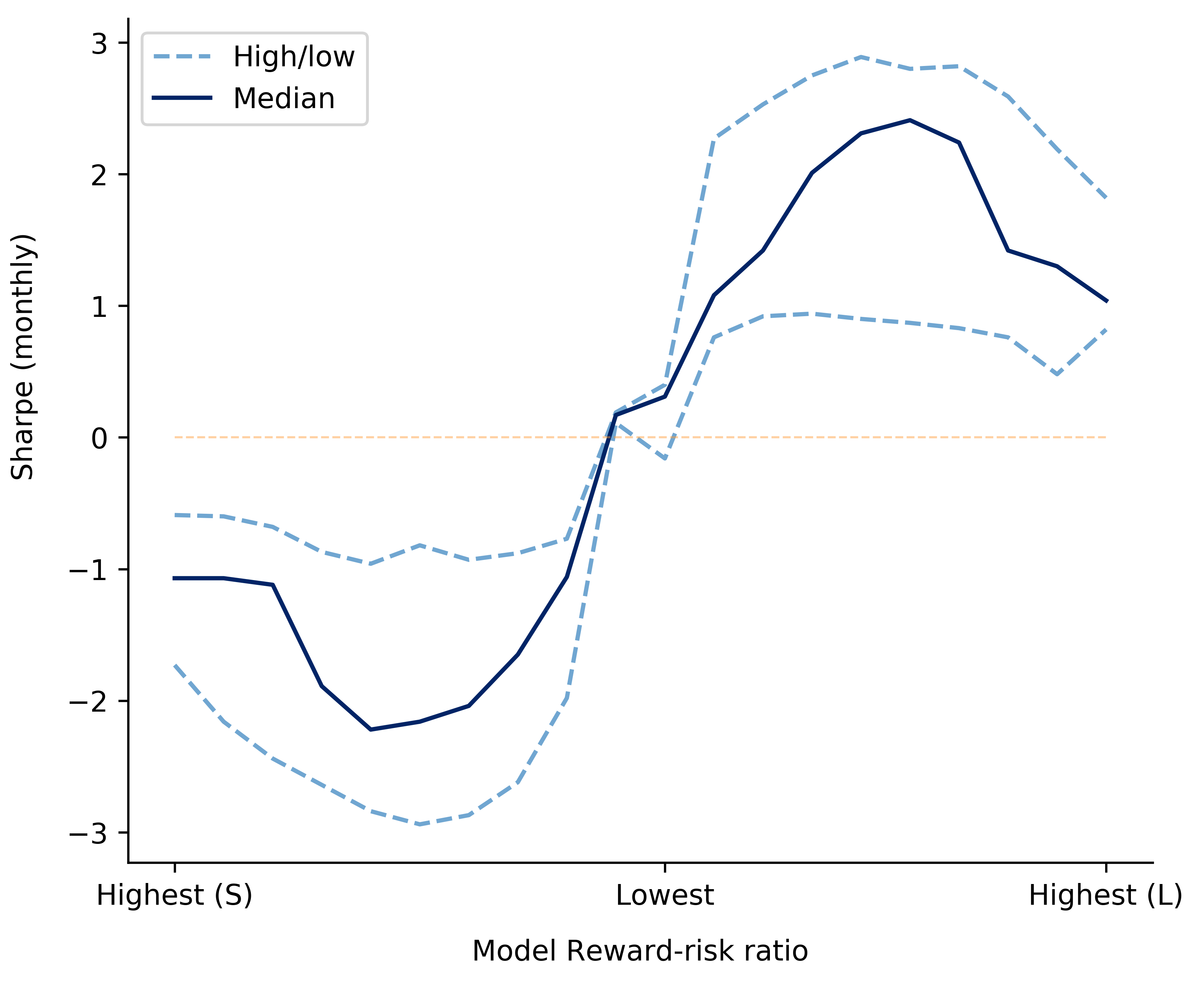}
  
\end{minipage}
\captionof{figure}{Left: Monthly Sharpe performance for baseline Bayesian OLS model, in analogy with Table 3, for estimating the full covariance matrix.  While the base strategy appears to show comparable performance to the more sophisticated models, it does not appear to generally improve taking uncertainty into account. Right:  Median out-of-sample model predicted risk-reward ratio (for months 8 to 12 of out-of-sample data) versus monthly Sharpe, for a diagonal covariance MLP.  The greatest long (short) ratio is shown to the right (left).  Here, out-of-sample Sharpe performance for the short positions is multiplied by $-1$.}
\label{fg:F}
\end{table}

\end{document}